\documentclass[twocolumn,aps,]{openjournal}
\usepackage{lineno,hyperref}
\modulolinenumbers[5]
\usepackage{ulem}

\usepackage{amsthm}
\usepackage{amsmath}
\usepackage{amssymb}
\usepackage{xcolor}
\usepackage{cpplistings}
\usepackage{xspace}
\usepackage{microtype}

\renewcommand{\cppfgcolor}{\color{Black}}

\lstdefinestyle{cpplistingstyle}
  { 
  , basicstyle = \ttfamily\small
  , breaklines          = false
  , emptylines          = 1
  , escapeinside        = {(*@}{@*)}
  , frame               = single
  , framexleftmargin    = .35em
  , numberblanklines    = false
  , numbersep           = 10pt
  , numberstyle         = \tiny
  , numbers             = none
  , tabsize             = 2
  , xleftmargin         = 1em
  }

\lstdefinestyle{cppstyle}
  { basicstyle          = \ttfamily
  , commentstyle        = \itshape
  , keywordstyle        = 
  , showstringspaces    = false
  }

\newcommand{\eg}{\textit{e.g.}}

\usepackage{outlines}
\usepackage{enumitem}
\usepackage{comment}

\newcommand{\lbin}{\Delta\lambda_i}
\newcommand{\zbin}{\Delta z_j}

\newcommand{\zt}{z^{\text{true}}}

\newcommand{\zo}{z^{\text{ob}}}
\newcommand{\lt}{\lambda^{\text{true}}}
\newcommand{\lc}{\lambda^{\text{cen}}}
\newcommand{\lo}{\lambda^{\text{ob}}}
\newcommand{\Rm}{R_{\text{cen}}}
\newcommand{\fc}{f_{\text{cen}}}

\definecolor{britishracinggreen}{rgb}{0.0, 0.42, 0.24}
\newcommand{\jim}[1]{\textcolor{britishracinggreen}{[{\bf JTA}: #1]}}

\newcommand{\datablock}{\cppcode{DataBlock}\xspace}
\newcommand{\csi}{\cppcode{CosmoSISScalarIntegrationModule}}
\newcommand{\cpp}{C\kern-0.15ex{+}\kern-0.1ex{+}\xspace}

\usepackage{eso-pic}

\AddToShipoutPictureBG*{%
  \AtPageUpperLeft{%
    \hspace{0.75\paperwidth}%
    \raisebox{-9\baselineskip}{%
      \makebox[0pt][l]{\textnormal{DES-2023-0755}}
}}}%

\AddToShipoutPictureBG*{%
  \AtPageUpperLeft{%
    \hspace{0.75\paperwidth}%
    \raisebox{-10\baselineskip}{%
      \makebox[0pt][l]{\textnormal{FERMILAB-PUB-23-481-PPD}}
}}}%

\begin{document}


 \title{Building an Efficient Cluster Cosmology Software Package for Modeling Cluster Counts and Lensing}



\author{
M.~Aguena$^{1}$,
O.~Alves$^{2}$,
J.~Annis$^{3}$,
D.~Bacon$^{4}$,
S.~Bocquet$^{5}$,
D.~Brooks$^{6}$,
A.~Carnero~Rosell$^{1,7,8}$,
C.~Chang$^{9,10}$,
M.~Costanzi$^{11,12,13}$,
C.~Coviello$^{3}$,
L.~N.~da Costa$^{1}$,
T.~M.~Davis$^{14}$,
J.~De~Vicente$^{15}$,
H.~T.~Diehl$^{3}$,
P.~Doel$^{6}$,
J.~Esteves$^{2}$,
S.~Everett$^{16}$,
I.~Ferrero$^{17}$,
A.~Fert\'e$^{18}$,
D.~Friedel$^{19}$,
J.~Frieman$^{9,3}$,
M.~Gatti$^{20}$,
G.~Giannini$^{21}$,
D.~Gruen$^{5}$,
R.~A.~Gruendl$^{22,19}$,
G.~Gutierrez$^{3}$,
K.~Herner$^{3}$,
S.~R.~Hinton$^{14}$,
D.~L.~Hollowood$^{23}$,
K.~Honscheid$^{24,25}$,
D.~J.~James$^{26}$,
T.~Jeltema$^{23}$,
M.~Kirby$^{27}$,
K.~Kuehn$^{28,29}$,
O.~Lahav$^{6}$,
P.~Li$^{30}$,
J.~L.~Marshall$^{31}$,
T.~McClintock$^{27}$,
D.~Mellor$^{32}$,
J. Mena-Fern{\'a}ndez$^{15}$,
R.~Miquel$^{33,21}$,
J.~O'Donnell$^{23,9,3}$,
A.~Palmese$^{30}$,
M.~Paterno$^{3}$,
M.~E.~S.~Pereira$^{34}$,
A.~Pieres$^{1,35}$,
A.~A.~Plazas~Malag\'on$^{36,18}$,
M.~Rodriguez-Monroy$^{15}$,
A.~K.~Romer$^{37}$,
A.~Roodman$^{18,36}$,
E.~Sanchez$^{15}$,
M.~Schubnell$^{2}$,
I.~Sevilla-Noarbe$^{15}$,
T.~Shin$^{38}$,
M.~Smith$^{39}$,
E.~Suchyta$^{40}$,
M.~E.~C.~Swanson$^{41}$,
G.~Tarle$^{2}$,
J.~Weller$^{42,43}$,
P.~Wiseman$^{39}$,
H.-Y.~Wu$^{44}$,
Y.~Zhang$^{45}$,
C.~Zhou$^{23,46}$ \\
 (DES Collaboration)
}

\begin{abstract}
We introduce a software suite developed for galaxy cluster cosmological analysis with the Dark Energy Survey Data. Cosmological analyses based on galaxy cluster number counts and weak-lensing measurements need efficient software infrastructure to explore an increasingly large parameter space, and account for various cosmological and astrophysical effects. Our software package is designed to model the cluster observables in a wide-field optical survey, including galaxy cluster counts, their averaged weak-lensing masses, or the cluster's averaged weak-lensing radial signals. To ensure maximum efficiency, this software package is  developed in C++ in the CosmoSIS software framework, making use of the CUBA integration library. We also implement a testing and validation scheme to ensure the quality of the package. We demonstrate the effectiveness of this development by applying the software to the Dark Energy Survey Year 1 galaxy cluster cosmological data sets, and acquired cosmological constraints that are consistent with the fiducial Dark Energy Survey analysis. 
\end{abstract}





\section{Introduction}


The abundance of galaxy clusters, the most massive gravitationally-bound structures in the Universe, are important cosmology probes utilized by wide-field cosmic surveys like the Dark Energy Survey (DES) \citep{2020PhRvD.102b3509A, 2021PhRvL.126n1301T}. Together with large scale structure correlation functions \citep{2008ARA&A..46..385F,2013PhR...530...87W}, galaxy cluster cosmology studies are projected to yield significantly improved constraints on dark energy in coming years, especially with the Rubin Observatory Legacy Survey of Space and Time (LSST) \citep{2018arXiv180901669T}. 

To derive cosmological constraints from galaxy cluster abundance observables \citep[see a review in][]{doi:10.1146/annurev-astro-081710-102514}, studies rely on a few theoretical ingredients to model the observations. First, gravity theories predict that galaxy clusters form through linear gravitational accretion and nonlinear collapse \citep[see a review in][]{2012ARA&A..50..353K}. Different cosmology models predict different abundances of galaxy clusters, and those abundances can be modeled with a dark matter halo mass function \citep[e.g.,][]{1974ApJ...187..425P, 2008ApJ...688..709T, 2019ApJ...872...53M}. This mass function depends on cosmological parameters such as the matter fluctuation amplitude and its growth, $\sigma_8$, and the total matter density, $\Omega_m$, and is subject to effects such as the number of neutrino species and the neutrino matter densities \citep{2013JCAP...12..012C}, as well as baryonic effects at the small scales \citep{2016MNRAS.456.2361B}. Further, the cosmic geometry predicted by cosmological models allows us to translate a dark matter halo mass function into dark matter halo counts \citep{2013PhR...530...87W}.  

To compare the abundance of dark matter halos to galaxy cluster observations, we also need to forward model how imaging surveys identify and measure galaxy clusters using a set of observational criteria \citep[e.g.,][]{2014ApJ...785..104R, 2016ApJS..224....1R, 2020MNRAS.493.4591P, 2021MNRAS.502.4435A}. 
The key to this task is a cluster observable-mass relation which models the relation between a cluster mass ``proxy" and the dark matter halo mass. This relation takes many forms in literature -- the Log-Normal relation is typical, 
but new forms of the relation have been suggested in recent years \citep[e.g.,][]{2015ApJ...810...21M, 2019MNRAS.482..490C, 2022arXiv220102167Z}.

Parameter inference in cluster cosmology is often done by adopting a Bayesian framework comparing theoretical predictions to observations. The posterior distributions of the cosmological parameters and nuisance parameters are often sampled using Markov Chain Monte Carlo (MCMC) methods, with a typical analysis requiring hundreds to thousands of CPU hours \citep{2022arXiv220208233L}. 
Additional systematic effects, such as incorrectly choosing a cluster's central galaxy known as ``mis-centering'' \citep{2019MNRAS.487.2578Z}, the projection of correlated and uncorrelated structures around the clusters \citep{2019MNRAS.482..490C}, the weak lensing cluster member contamination   \citep{2019MNRAS.489.2511V}, and the shape and orientation of real non-spherical clusters also need to be considered as corrections to the theoretical
expectation from spherically symmetric cluster density distributions.   
Given all of those elements, forward-modeling the galaxy cluster abundance and their shear measurements can be approached as an integration problem because of the need to marginalize over various theoretical and observational effects. Solving this problem poses a significant computational challenge when the integrations have large dimensions and need to be computed at every step of a Monte Carlo parameter sampling analysis.

In this paper, we introduce our efforts to develop a software package to compute models for cluster abundance and weak lensing  measurements. The latter introduces a few more dimensions of integrations than the fiducial analysis adopted in DES Y1 \citep{2020PhRvD.102b3509A}. 
As the MCMC sampling methods used for cosmological parameter inference often need to compute those models with varying cosmological and nuisance parameters for many thousands of times, the efficiency of those computations becomes critically important. The change necessitates the development of a new pipeline that is focused on computing efficiency and the application on high-performance computing resources. In the rest of this paper, we describe the design background of this new pipeline including the theoretical models,  the structure of our software package and our testing mechanisms. We also use this new package to reproduce the fiducial DES Y1 analysis in Section ~\ref{sec:analysis} and acquire consistent cosmological results. This analysis also serve  as an independent check of the fiducial DES Y1 analysis. 

{The package described in this paper will be made openly available}.

\section{Cluster Cosmology Theoretical Framework}\label{s:theory}

\subsection{Number Counts and Average Masses}

In galaxy cluster cosmology studies performed by optical surveys like Sloan Digital Sky Survey (SDSS) and DES, the clusters are discovered through their richness, denoted as $\lambda$ \citep{2012ApJ...746..178R}. 
To-date, the cluster cosmology analyses performed with large cluster samples from SDSS and DES \citep{2010ApJ...708..645R, 2019MNRAS.488.4779C, 2020PhRvD.102b3509A} use summary statistics of the selected samples, i.e. the number of clusters in a few richness and redshift ranges, and their average masses in these richness/redshift ranges as the observable of cosmology. In this section, we describe the analytical formalism for modeling those cluster number and mass observables. 

We first present the theoretical predictions for the total number and the average mass of the clusters in one redshift and richness bin. Those quantities can be written as 
\begin{equation}
\begin{split}
  \left<N(\lbin,\zbin)\right> =  
    &\int_{\zbin} d\zo  \int_{\lbin} d\lo \int_0^\infty dM \int_0^\infty d\zt  \\ 
  & P(\zo|\zt,\lbin) P(\lo|M,\zt) \\
  & n(M,\zt)\ \Omega(\zt)\frac{dV}{d\Omega\,d\zt}(\zt)  \ . \\
   \left<NM(\lbin,\zbin)\right> =  
      &\int_{\zbin} d\zo \int_{\lbin} d\lo \int_0^\infty dM \int_0^\infty d\zt  \\ 
    &  P(\zo|\zt,\lbin) P(\lo|M,\zt) \\
    & M\cdot n(M,\zt)\  \Omega(\zt)  \frac{dV}{d\Omega\,d\zt}(\zt) \ . \\
\left<M(\lbin,\zbin)\right> = & \left<NM(\lbin,\zbin)\right>/\left<N(\lbin,\zbin)\right>.
\end{split}
\label{eq:1}
\end{equation}
In this set of equations, $\left<N(\lbin,\zbin)\right>$ is the predicted number of galaxy clusters in a redshift range  of $\zbin$ and a richness range of $\lbin$, while $\left<NM(\lbin,\zbin)\right>$ is the predicted total mass of those clusters and $\left<M(\lbin,\zbin)\right>$ is their average mass. 

The predictions for those three quantities consider several theoretical elements. First, the volume density of galaxy clusters in terms of their theoretical quantities, mass $M$ and redshift $\zt$, is known as the halo mass functions $n(M, \zt)$ . Analytical approximations for the halo mass function can be calculated from the matter power spectrum and the cosmological model, and depends most strongly on the values of $\Omega_m$ and $\sigma_8$ in a $\mathrm{\Lambda CDM}$ cosmology \citep[our applications are based upon][]{2008ApJ...688..709T}.  Second, $\Omega(\zt)$ is the total solid angle area of the cosmic survey, which may have a redshift dependence as  the usable area for cluster searching changes with redshift.  Further, $\frac{dV}{d\Omega\,d\zt}(\zt)$ is the cosmological volume element in solid angle $d\Omega$ and a redshift slice $d\zt$, which also depends on the cosmological parameters.  On top of those cosmology and geometry related quantities, we also consider a few relations linking cluster observable to their theoretical quantities. Among those, $P(\zo|\zt,\lbin)$ considers the redshift uncertainty of the clusters, as its measured value of $\zo$ may deviate from its true redshift characterized by a (sometimes Gaussian) probabilistic distribution. $P(\lambda|M,\zt)$ relates the cluster's observational selection quantity $\lambda$  to the cluster's theoretical mass quantity, $M$, and true redshift $\zt$. $P(\lambda|M,\zt)$ can be further expanded to marginalize over latent parameters to consider astrophysical effects like projection and cluster triaxial shapes. A more detailed description can also be found in \cite{2020PhRvD.102b3509A}.

\subsection{Number Counts and Average Lensing Signals}

In observations, to measure the averaged cluster masses, the averaged cluster lensing signals are analyzed for the clusters in the richness ranges \citep{2019MNRAS.482.1352M}. In this software development, our end goal is to forward model this lensing observable together with the cluster number counts, instead of relying on the cluster's average mass measurements and number counts. Realizing this goal requires us to further consider the relations between cluster mass, concentration ($c$), and the cluster's projected radial mass density from the center $\Sigma_{cen}(r|M, \zt, c)$. 

\begin{align}\label{eq-centered-shear-profile}
\left<N \Sigma_\mathrm{cen}(r, \lbin,\zbin, c)\right> = \hspace{-10em} &\\ \nonumber
   & \int_{\zbin}d\zo \int_{\lbin} d\lambda\int_0^\infty dM \int_0^\infty d\zt \\  \nonumber  
   &  P(\zo|\zt,\lbin) P(\lambda|M,\zt)
     \\ \nonumber
& \Sigma_{cen}(r|M, \zt, c)  \cdot n(M,\zt)\ \Omega(\zt)\frac{dV}{d\Omega d\zt}(\zt) \\ \nonumber
\end{align}

Further, those projected mass densities are susceptible to an additional physical effect -- mis-identification of galaxy cluster centers, that the cluster's real centers can be offset from the identified centers by a distance, $\Rm$. This mis-centering offset can be described by a probability distribution $P(\Rm)$:
\begin{equation}\label{eq:roffset}
\begin{split}
&P(\Rm) =\fc \delta(\Rm)+(1-\fc)P_{mis}(\Rm). \\
\end{split}
\end{equation}

In this equation, we recognize that a significant fraction ($\fc$) of the clusters are actually well-centered. For those well-centered clusters, $\Rm$ can be described as a $\delta$ function, as often adopted in cluster lensing modeling \citep[e.g.,][]{2019MNRAS.482.1352M}. For the clusters that are indeed mis-centered, their mis-centering distance can be described by a continuous probabilistic function $P_{mis}(\Rm)$, constrained by multi-wavelength studies \citep[e.g.,][]{2019MNRAS.487.2578Z}.  Further, mis-identification of cluster centers have also been shown to further affect the cluster richness observable by introducing dependence on the cluster's mis-centering offset, which can be modeled by an additional latent parameter $\lc$ describing a cluster's would-be richness without the mis-centering effect \citep{2007ApJ...656...27J, 2019MNRAS.487.2578Z}.
\begin{equation}
\begin{split}
   & P(\lo|M, \zt,\Rm)= \\
   &\int_0^\infty d\lc\, P(\lo|\lc, \Rm)P(\lc|M, \zt).
   \end{split}
\end{equation}
This equation marginalizes over the latent $\lc$ parameter, assuming that the cluster's observed richness $\lo$ depends on $\lc$ and the cluster's mis-centering offset $\Rm$ with a probability distribution $P(\lo|\lc, \Rm)$.

Thus, the prediction for the clusters' projected surface mass density as a function of radius can be written as, 
\begin{align}\label{eq:shear-profile-BREAKDOWN}
\left< \Sigma(r|\lbin,\zbin, c) \right> =  \hspace{-4em}& \\ \nonumber
   & \frac{\fc}{\left<N(\lbin,\zbin)\right>} \left<N \Sigma_\mathrm{cen}(r, \lbin,\zbin, c)\right> \ + \\ \nonumber
   & \frac{1-\fc}{\left<N(\lbin,\zbin)\right>} \left<N \Sigma_\mathrm{mis}(r, \lbin,\zbin, c)\right> \ . 
\end{align}
While $\left<N \Sigma_\mathrm{cen}(r, \lbin,\zbin, c)\right>$ has been defined in Equation\ref{eq-centered-shear-profile}, $\left<N \Sigma_\mathrm{mis}(r, \lbin,\zbin, c)\right>$ needs to azimuthally average over the mis-centering distance distribution, which can be written as:
\begin{align}
\left<N\Sigma_\mathrm{mis}(r, \lbin,\zbin, c)\right> =  \hspace{-8em} &\\ \nonumber
   & \int_{\zbin}d\zo \int_{\lbin} d\lo \int_0^\infty dM \int_0^\infty d\zt \\ \nonumber
   & \int_0^\infty d\lc \int_0^\infty d\Rm \int_{0}^{2\pi} \frac{d\alpha}{2\pi} \\ \nonumber
& P(\zo|\zt,\lbin)\ P(\lo|\lc,\Rm)\ \\
& P(\lc|M,\zt)\ P_{mis}(\Rm)   \\ \nonumber
   & n(M,\zt)\ \Omega(\zt)\frac{dV}{d\Omega d\zt}(\zt) \\ \nonumber
   & \Sigma_{cen}(\sqrt{r^2+\Rm^2-2r\Rm. \mathrm{cos}\alpha}|M, \zt, c) .
\end{align}

Given that mis-centering affects the cluster richness-mass relation, we shall also update the writing of the cluster number counts as 

\begin{align}\label{eq:abundance-BREAKDOWN}
\left<N(\lbin,\zbin)\right> =\   &\fc \cdot  \left<N_\mathrm{cen}(\lbin,\zbin)\right>&\\ \nonumber
    &+  (1-\fc) \left<N_\mathrm{mis}(\lbin,\zbin)\right>, \\ \nonumber
 \left<N_\mathrm{cen}(\lbin,\zbin) \right> & =    \\ \nonumber
   & \hspace{-4em} \int_{\zbin} d\zo \int_{\lbin} d\lo \int_0^\infty   dM \int_0^\infty d\zt  \\  \nonumber
   & \hspace{-4em} P(\zo|\zt,\lbin)\ P(\lo|M,\zt)  \\ \nonumber
   &  \hspace{-4em} n(M,\zt) \ \Omega(\zt)\frac{dV}{d\Omega\,d\zt}(\zt) , \\ \nonumber
\left<N_\mathrm{mis}(\lbin,\zbin)\right> &=  \\ \nonumber
   & \hspace{-4em} \int_{\zbin} d\zo \int_{\lbin} d\lo  \int_0^\infty dM \int_0^\infty d\zt \\ \nonumber
   & \hspace{-4em} \int_0^\infty d\lc \int_0^\infty d\Rm  \\  \nonumber
   & \hspace{-4em} P(\zo|\zt,\lbin)\ P(\lo|\lc,\Rm)\ \\ \nonumber
   & \hspace{-4em} P(\lc|M,\zt)\ P_{mis}(\Rm)   \\ \nonumber
   &  \hspace{-4em} n(M,\zt) \ \Omega(\zt)\frac{dV}{d\Omega\,d\zt}(\zt) \ .
\end{align}

Finally, the results of the cluster's averaged projected densities can be further converted into the so-called excess surface mass density $\Delta\Sigma$ as with:
\begin{multline} \label{eq:DS-breakdown}
\left< \Delta \Sigma(R|\lbin,\zbin, c) \right>  = \\ \frac{2}{R^2}\int_0^R r \left< \Sigma(r|\lbin,\zbin, c) \right> \mathrm{d} r -\left< \Sigma(R|\lbin,\zbin, c) \right>.
\end{multline}
or into reduced tangential shear signals, $\gamma_t$, which are observables from cluster weak lensing analyses \citep{2020A&ARv..28....7U}.

\section{Parameter Estimation and CosmoSIS}
\label{sec:param_est}

\begin{figure*}
  \centering
  	\includegraphics[width=160mm]{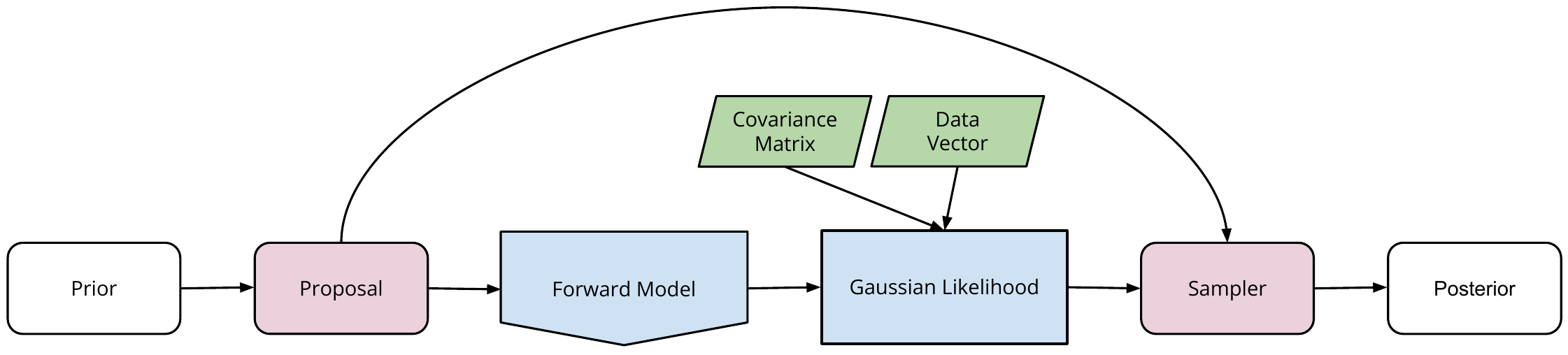}
  \caption{Components of CosmoSIS pipeline using a Markov Chain Monte Carlo sampling method to derive the posterior cosmological constraints.}
  \label{fig:cosmosis_flowchart_samp}  
\end{figure*}

Our development is guided by the Bayes' theorem to derive a posterior probability distribution of a set of model parameters $\theta$, upon a set of observations $x$. 
\begin{equation}\label{eq:jointposterior} 
  P(\theta \, | \, x, I) \propto P(x \, | \, \theta, I) \, P(\theta \, | \, I) , 
\end{equation}
In this equation, our knowledge of the parameters before consideration of the data $x$, but based on our prior information $I$, is described by the \emph{joint prior probability} $P(\theta \, |\, I)$. On the other hand, $P(x \, | \, \theta, I)$, often called the \emph{likelihood}, is the probability of observing the data $x$, given the parameters $\theta$ and the same prior information $I$.

In cluster cosmology studies that rely on summary statistics, such as the cluster number counts and their lensing masses or their average lensing profiles described in the previous section, the likelihood is often assumed to be a multi-dimensional Gaussian as studied and tested in literature \citep{2015MNRAS.449.4264G, 2019MNRAS.490.2606W, 2021A&A...652A..21F}. The mean of the Gaussian is the expectation values of the predictions for the data in a given cosmology ($\mu(\theta)$) while the spread of the distribution is described by a covariance matrix $C$ that characterizes the uncertainties in both observation and theory. Thus the likelihood function can be written as:  
\begin{equation}\label{eq:likelihood} 
  P(x \, | \, \theta, I) = \mathcal{N}(x-\mu(\theta), C(\theta))
\end{equation}

To derive the posterior cosmological constraints given the observational measurements, theoretical predictions, and the covariance matrix, we rely on MCMC methods, as well as other sampling techniques. These samplings and the development of the prediction code are performed in the  CosmoSIS software framework \footnote{https://bitbucket.org/joezuntz/cosmosis}. 
\autoref{fig:cosmosis_flowchart_samp} demonstrates a typical workflow to derive posterior cosmological constraints in CosmoSIS.


CosmoSIS \cite{zuntz:2015} is a modular system for cosmological parameter estimation. 
 It solves integrals like that in \autoref{eq:jointposterior} using MCMC. MCMC solves an $n$-dimensional integral by generating a collection of points, called \emph{samples}, in an $n$-dimensional space; the points are distributed according to the functional form of the solution.
A \emph{sampler} drives the exploration of the parameter space. A sampler is the software component responsible for determining the location in the parameter space for the next step of the MCMC chain based upon the evaluation of the user-supplied likelihood function, and the location of the current sample. The user provides the likelihood function by supplying one or more modules (\emph{Forward Model} modules and/or \emph{ Gaussian Likelihood} modules in \autoref{fig:cosmosis_flowchart_samp}) which, given a point in the parameter space, evaluate the likelihood of the data given those parameters.

In CosmoSIS, a \emph{module} is user-written code responsible for calculating physical quantities of interest (\eg, predicted cluster number counts) based on the parameters of the current sample.
CosmoSIS modules can be written in several programming languages, including Python, C, Fortran, C++ and Julia. Calculations from those modules are stored in a structure known as a \datablock object, which is passed through the list of modules used in a CosmoSIS pipeline setup, and thus functions to pass results between modules.

\section{Pipeline structure}


Our pipeline delivers prediction code that can be implemented in a CosmoSIS \emph{module} for the likelihood calculation (Equation~\ref{eq:likelihood}), $\mu(\theta)$. In particular, we calculate the predictions for the cluster number counts, cluster masses and/or shear profiles using models and the integrals they involve. The efficiency of the numerical integration library we use is of primary importance, as the calculation of the integrals of interest can involve $10^5$---$10^6$ function evaluations. In the rest of the section, we describe the structures of this pipeline as well as how those structures function in an  analysis implementation.


\subsection{``Models"}\label{ss:models}

The smallest software units in our development of the pipeline are the \emph{model}s. Examples  
include the survey area function $\Omega(\zt)$, 
the halo mass function $n(M,\zt)$, 
and the cluster observable-mass relation $P(\lambda|M,\zt)$ from \autoref{eq:1}. 

In each case, the essence of one \emph{model} is a function which we need to evaluate as part of our integrand, and which can be tested for correctness in isolation from the rest of the integrand. Furthermore, there may be more than one sensible implementation of the function---for example, in the case of the survey area function, the implementation would be different for different surveys. Finally, the evaluation of the function may require the use of some additional data, other than the values of the function arguments. A very common example of this is the use of interpolation tables, which can be output arrays from other CosmoSIS modules (such as the CosmoSIS CAMB module), or derivations from supporting analyses not as part of this efforts.

Figure~\ref{lst:omega_sdss} shows most of the implementation of one of our \emph{model}s, the survey area for the SDSS redMaPPer cluster sample, which is redshift dependent. The name of the class is \cppcode{OMEGA_Z_SDSS}. The state of each object of this type consists only of a \cppcode{polynomial} object named \cppcode{fit_}, which itself is a callable object representing a polynomial interpolation function. The \emph{default constructor} of the class, which initializes an object of a class, establishes the state of the object, in this case a polynomial used to evaluate the function. There is a second constructor, which takes a CosmoSIS \datablock object; this use of the constructor will be described later. In this case, the constructor does nothing differently than the default constructor---it is acceptable for a constructor to ignore its argument(s).  The function call operator shows that this class represents a function of one argument, \cppcode{zt}, and the implementation is using the interpolation table to calculate the function. Importantly, the function call operator is implemented \emph{inline}. This means the compiler is free to put the code of the function directly into the place where the function is called. This is important because it provides the compiler the opportunity to see the implementation of the function at any place in the code where this function is called, which allows an optimizing compiler to make use of this information to do more optimizations than would be possible without this information. This member function is \cppcode{const}-qualified, which means calling the function does not alter the state of the object. Marking the function helps assure the correctness of our code; if an error in implementation is made that perhaps inadvertently modifies the state of the object (for example, by changing the values in an interpolation table), a compilation failure will occur. Marking the function \cppcode{const} can also provide a compiler additional opportunities for optimization of the generated machine code.

The example class shown in Figure~\ref{lst:omega_sdss} computes the survey area for SDSS, but additional classes/\emph{model}s that represent the survey area of different data sets, e.g., DES year 1 data sets, or year 6 data sets can be constructed.
Each class that represents a survey area has a function call operator with the same signature: one argument, of type \cppcode{double} representing the redshift at which the area element is to be calculated, and returning a \cppcode{double} representing the area of the survey at that redshift.

In addition, every \emph{model} (not just those representing cosmological area calculations) has a constructor that takes a CosmoSIS \datablock object. This constructor is used when the \emph{model} is invoked in a CosmoSIS pipeline, and is used to provide any necessary initialization values to the \emph{model}. In this case, no such initialization parameters are needed, and this constructor does not need to consult the \datablock  which is given. However, the constructor must still be present, so that the generic code that creates a \emph{model} (and which provides the \datablock to be used) is well-formed. Our library of models includes two volume elements, one for the DES and one for the SDSS.

\begin{figure}
\begin{cpplisting}
class OMEGA_Z_SDSS {
private:
  polynomial<12> fit_{...};

public:
  OMEGA_Z_SDSS() = default;
 
  OMEGA_Z_SDSS(cosmosis::DataBlock&) {}

  double
  operator()(double zt) const
  {
    return fit_(zt - 0.2);
  }
};
\end{cpplisting}
\caption{Abridged implementation of the class \cppcode{OMEGA_Z_SDSS}, which represents the cosmological survey area for SDSS. Details of namespaces and the implementation of the initialization of \cppcode{fit_} have been suppressed for brevity.}
\label{lst:omega_sdss}
\end{figure}

We also require that those \emph{model}s, as they are the smallest development unit, must have a corresponding test whenever possible. This is described in more detail in \autoref{ss:unittests}.

\subsection{CosmoSIS modules in general}


The models are embedded into CosmoSIS modules.
The module interface provided by CosmoSIS has a natural implementation when translated into \cpp. Each module is represented as an instance of a \emph{module class}. We note that no inheritance is necessary, because there is nothing in CosmoSIS that would ever interact with the base class of an inheritance hierarchy.  

\begin{figure}

  \centering
  	\includegraphics[width=55mm]{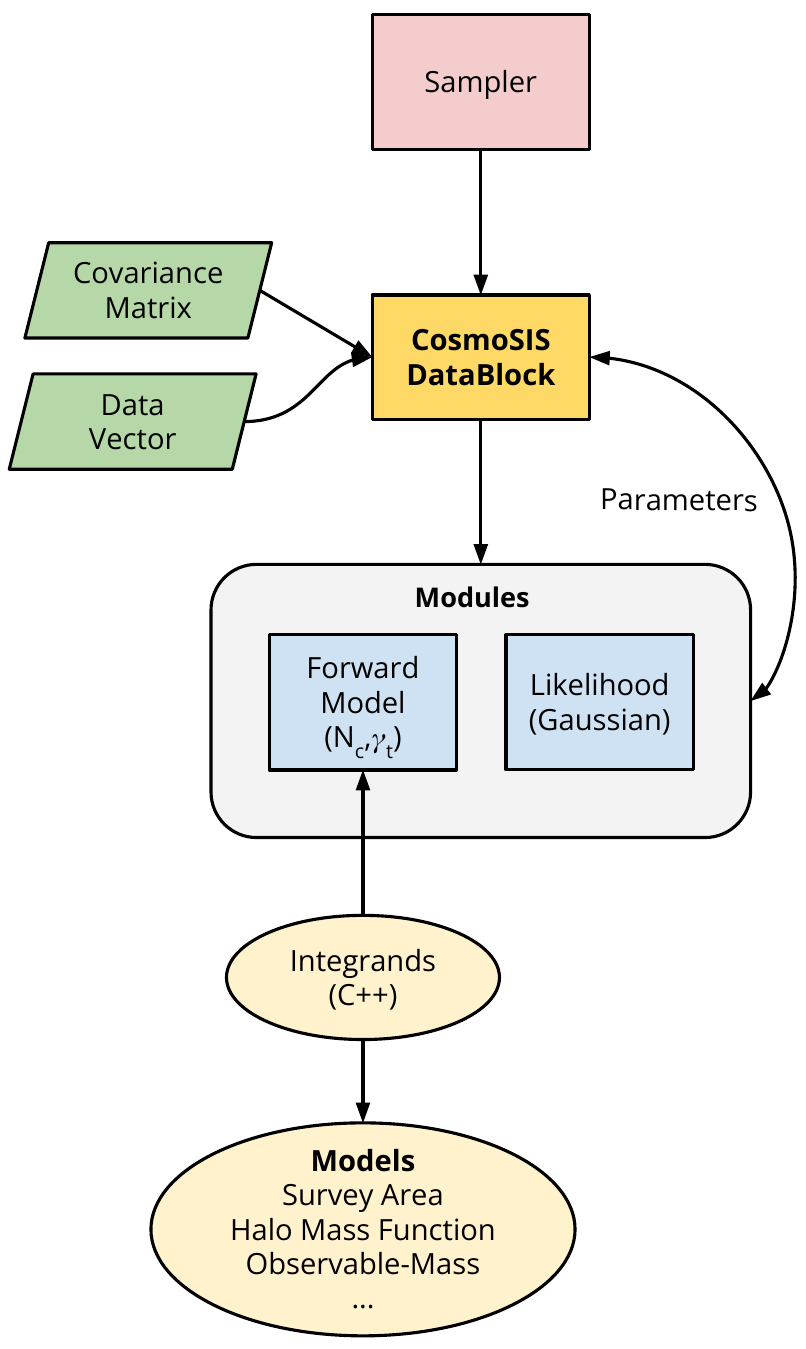}
  \caption{Flowchart of the CosmoSIS cluster pipeline summarizing the operations between integrands, modules and \datablock.}
  \label{fig:cosmosis_flowchart_pipe}  
\end{figure}

A module object is created at program start-up time, using the constructor of the class. Any configurable parameters required by the class are passed to the constructor through a \datablock object, which contains the configuration information read by the CosmoSIS program. At each step of the MCMC process, CosmoSIS calls the module's \emph{execute} 
function; this is naturally implemented in \cpp as a member function \cppcode{execute}, which is passed an instance of the class \emph{DataBlock} carrying the cosmological parameters and derived physical quantities for the current MCMC sample. For our prediction modules, these quantities are used as inputs for calculating the predictions of observables, given the cosmology of the current sample. The prediction modules in turn record these predictions in the \datablock for later processing by our likelihood module. Our likelihood module retrieves the predictions from the \datablock and calculates the data likelihood for the sample cosmology, recording it in the \datablock. \autoref{fig:cosmosis_flowchart_pipe} shows a flowchart of the operations between integrands, modules and the \datablock described in this section. Finally, the module objects are destroyed at program shutdown. In a well-designed class, there is usually no user-written code necessary for the destructor; the compiler-generated destructor will assure that all allocated resources are properly released.

\subsection{Our physics modules}\label{sec:physics_module}
The prediction modules calculate their predictions based on integrands formed from the models described in \autoref{ss:models}. These models are typically created for a given cosmology, and thus model instances are created anew for each MCMC sample. (In some cases, as described below, models have an even shorter lifetime.) \autoref{eq-centered-shear-profile} shows one such integral, which calculates the centered shear profile. This integral is done over five variables of integration. Two of the variables of integration ($\zo$, $\lo$) are each integrated over several distinct, possibly overlapping, ranges. Thus there are distinct 5-dimensional volumes of integration to be calculated for each sample. In addition, the integral is evaluated for a distinct set of values of $r$ and $c$, which form a grid of locations in $(r, c)$ space at which each of the integrals are calculated. Typically, several hundred integrals are calculated for a single integrand---the product of the number of volumes of integration and the number of grid points at which the integral is evaluated.

\subsection{The physicist supplied integration problem class}

To simplify the task of writing a prediction module, we have created the class template \cppcode{CosmoSISScalarIntegrand} that takes a user provided template parameter and supplies the integrand. The template provides several facilities. It configures the model upon construction, when possible. It re-configures the model for each sample, if needed. It re-configures the model for the grid points at which the integrands are integrated. It configures the volumes of the integrand, and calculates all the integration results. It then organizes the integration results into simple data structures and puts them into the \cppcode{DataBlock}. The type used for this template argument must be a class (or struct).

We have developed a class template (\csi) that employs these tools to make the writing of a physics module that performs integrals like those described above simpler.
The physicist who wishes to write an integration module can do so with the help of this template, and then needs to write a simpler class (named \cppcode{X}, for example) that provides the following features; the type \cppcode{CosmoSISScalarIntegrationModule<X>} is then a fully-functional CosmoSIS module.

The required features of the physicist-provided class include not only the implementation of the function to be integrated (which is implemented through the function call operator of the class), but also functions that provide the information needed to make the module configuration code function for the specific integrand.

The physicist inventing a new module must write a class with several features that are relied upon by the class template \csi.
These features include some required nested types (type definitions within the class), static and non-static member functions, and static member data.
In the remainder of this subsection we describe these features.

The class must have two nested type declarations, \cppcode{grid_t} and \cppcode{volume_t}.
The type \cppcode{grid_t} is to be defined using the helper template \cppcode{y3_cluster::grid_t<M>}, using the appropriate value of \cppcode{M} for the dimensionality of the desired grid points (as described in Section~\ref{sec:physics_module}, grid points are set of variable values at which the integrals are evaluated).
The type \cppcode{volume_t} is to be defined using the helper template \cppcode{cubacpp::IntegrationVolume<N>}, and using the appropriate value of \cppcode{N} to define the dimensionality of the volume of integration (otherwise also known as the  boundaries of the parameters to be integrated).

The class must have a constructor that accepts a single \datablock object.
This object carries the configuration parameters provided by the CosmoSIS user's configuration file.
Parameters that do not change with MCMC sample or with the grid point at which the integral is being evaluated should be set in this constructor.

The class must provide a function call operator that defines the actual integrand.
This must define a (non-static) member function with a return type of \cppcode{double} and which accepts any fixed number of \cppcode{double} arguments.
At compilation time, this member function will be probed to determine the dimensionality of the resulting integrand, and to verify the dimensionality of the integration volume specified in the type \cppcode{volume_t}.

The class must provide a member function  \cppcode{set_sample} that accepts a \datablock object.
This function is called with every MCMC sample; the argument carries all the data (most importantly, the cosmology) corresponding to the current sample.
The integrand object can read whatever data are needed to adjust model parameters to reflect the current sample.

The class must provide a \cppcode{static} member function \cppcode{make_integration_volumes}, which accepts a \datablock.
This function is to be implemented using one of two provided helper function templates, \cppcode{make_integration_volumes_wall_of_numbers}  and \cppcode{make_integration_volumes_cartesian_product}.
The choice of helper used by the integrand class determines how the integration volumes used by resulting CosmoSIS module are configured.
Both helper templates are called with a number of C-style strings specifying the names of the parameters that are to be read from the CosmoSIS module configuration; the number of strings specified must match the declared dimensionality of the integrand's grid.
The compiler will reject as invalid code any class for which these two do not agree.
The values of these parameters read at the startup of a CosmoSIS run determines the set of points at which the integrals are to be evaluated.
Note that the author of the class determines the dimensionality of the integration volume(s) and the name(s) of the configuration parameter(s) to be read from the CosmoSIS initialization file, but the user of CosmoSIS determines the number of integration volumes to be used in the run of CosmoSIS, and their sizes.

The class must provide a non-static member function \cppcode{set_grid_point} that is called to allow the setting of any model parameters that define the grid of points at which the integral is calculated for each sample.
This function is passed an object of type \cppcode{grid_point_t}, carrying the data that specifies the location of the grid point at which the integrand will next be evaluated.
The \csi class template implements the loop over the grid points specified in the CosmoSIS user's configuration file.

\subsection{Using the module}

We rely upon several of the integration routines provided by the C-language library \texttt{CUBA}\footnote{http://www.feynarts.de/cuba/}\citep{2007CoPhC.176..712H}.
Primarily, we use the (deterministic) algorithm CUHRE, described in \citep{10.1145/210232.210233};
we also make some use of the Monte Carlo algorithm VEGAS \citep{1978JCoPh..27..192L}.
The \texttt{CUBA} library requires an inconvenient form for the functions it integrates, and also integrates only over the unit hypercube.
It is left to the \texttt{CUBA}  user to provide the necessary transformation of the integrand to handle the integration volume of interest to the user.
To solve these issues, we use the \texttt{cubacpp}\footnote{https://bitbucket.org/mpaterno/cubacpp} \cpp library.
This allows the user-supplied integrand to be any ``callable object'' that takes any number of \cppcode{double} arguments, and returns \cppcode{double}.
It also allows the user to supply an arbitrary volume of integration, and assures at compilation time that the volume of integration is of the same dimensionality as the integrand.
The user of the algorithm can specify both \emph{relative} and \emph{absolute} tolerances, and the integration algorithm stops when either condition is met (or when a specified maximum number of integrand evaluations has been done, providing a coarse means of limiting the time allowed to calculate an integral).
The value returned from an integration problem includes the estimate of the integral, and an approximate error in that estimate, as well as some diagnostic information, most importantly whether the integration algorithm has converged or not.

A physics module created using the class template described above, instantiated using a physicist-supplied integrand class, has several configuration parameters that are to be set by the user of CosmoSIS. The author of the integrand class has specified the mathematical models to be used, the dimensionality of the integrand in question, and the parameters that are used to define  the locations in parameter space where the integrals are to be calculated. The user of CosmoSIS, not the author of the integrand class, chooses how many grid locations are to be used, and where they are. The user of CosmoSIS also defines the set of volumes over which the integral should be evaluated. The user also chooses which of the available integration routines (CUHRE or VEGAS) should be used, and configures the relative and absolute error tolerances for the integration. The user also specifies a maximum number of function evaluations to be used for each integration task, as a means of limiting the running time of the module for cases when the algorithm chosen is not converging rapidly enough.


\section{Tests and Physics Validation}

To ensure a robust pipeline, we utilize both unit tests and physics-based validations. In this set-up, a unit test is designed to test the models or infrastructural functions (e.g., 1-d and 2-d interpolators), while in cases that a crucial calculation makes use of multiple models and/or infrastructural functions, we rely on a physics validation document to check the quality of the calculations.

\subsection{Unit Tests}\label{ss:unittests}

\begin{figure}
\begin{cpplisting}
TEST_CASE("omega_z_sdss works")
{
  // read in comparison data file
  std::ifstream infile = 
  std::vector<double> zs;
  std::vector<double> ys;
  while (infile) {
    double z, y;
    infile >> z >> y;
    zs.push_back(z);
    ys.push_back(y);
  }

  OMEGA_Z_SDSS omega;

  //Perform the unit test comparison 
  for (std::size_t i = 0, 
  sz = zs.size(); i != sz; ++i) {
  
    double const fz = omega(zs[i]);
    double constexpr epsrel = 1.0e-6;
    CHECK(fz == 
    Approx(ys[i]).epsilon(epsrel));
    
  }
}
\end{cpplisting}
\caption{Abridged Unit test code for \cppcode{OMEGA_Z_SDSS}, which computes the cosmological survey area for the redMaPPer cluster catalog from the Sloan Digital Sky Survey (SDSS).}
\label{lst:omega_sdss_test}
\end{figure}

We use unit tests to ensure the accuracies of the models in our pipeline. These unit tests are implemented and ran through the CTest tool \footnote{https://cmake.org/cmake/help/latest/manual/ctest.1.html}. An example of the unit test is shown in \autoref{lst:omega_sdss_test}. 

The structures of the unit tests are organized as the follows. A comparison data set is read in and the relevant function is called to execute at the comparison data points. The output from the function is compared to the values in the comparison data sets. If the two values are comparable within an acceptance limit for every comparison point, the test passes. Otherwise, the test fails. 

To set up the unit tests for the physical models, we acquire expected values of those model functions at a given set of input values from previously implemented code in other analysis, e.g., Fortran90 code implemented for SDSS cluster cosmology analysis \citep{2019MNRAS.488.4779C}, or Python code implemented for quantifying cluster mis-centering effects \citep{2019MNRAS.487.2578Z}. We require the model output and the comparison value sets to agree either within a relative or an absolute (in the case of values being zeros) accuracy limit. The relative accuracy limit, meaning the allowed relative differences between the values, are by default set at $10^{-6}$ while the absolute accuracy limit, which is the absolute differences between the values, is set at $10^{-12}$ by default. In the cases that the comparison values are calculated through approximations or interpolations, we relax the accuracy requirements on a case-by-case basis. 

These unit tests are designed to be re-ran with a \textsc{ctest} command when the code re-compiles because of updates or revision and give a pass or fail indication. Output from some of the unit tests, especially those implementing physics models, are saved and can be used for debugging or visualization later.

\subsection{Physics Validation}

\begin{figure*}
  \centering
  	\includegraphics[width=150mm]{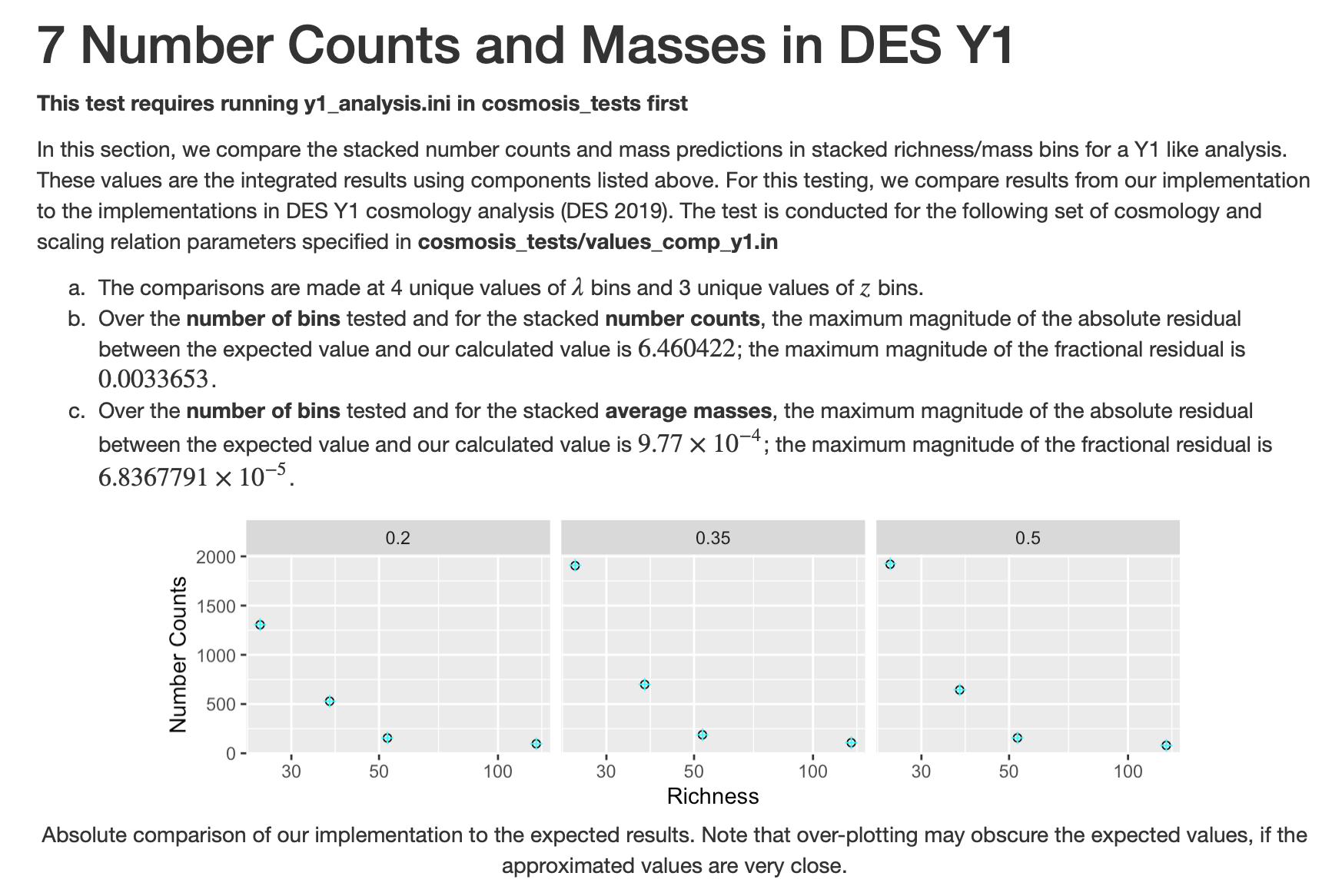}
  \caption{Excerpt of the physics validation output implemented in R. The example shows the predictions from our pipeline for the DES Y1 cluster number counts, compared to the model output in \citep{2020PhRvD.102b3509A} using the same set of values for cosmological and nuisance parameters. The two pipelines give highly consistent predictions. The plots and values of relative and absolute differences are automatically compiled from CosmoSIS output.}
  \label{fig:y1analysis_validation}
\end{figure*}

We also compile a physics validation document, which visualizes the unit test results and in addition,  visualizes a selected set of more complicated calculations from the physics modules. Specifically, we compare the module outputs that calculate the theoretical predictions for a set of SDSS/DES cluster observables, to the theoretical outputs calculated by previous analyses \citep{2019MNRAS.488.4779C, 2020PhRvD.102b3509A}. 

In designing the physics validation document, our original goal was to implement the validation in a system that enables version control to easily track  the comparisons. We have considered a few different systems including static pages such as word or PDF. Jupyter notebooks are also considered, but ruled out because they are saved in XML which makes it difficult to interpret the version differences. In the end, the validation is implemented as R markdown notebooks, which makes it easier to track versions and present the differences. 

 The physics validation R notebooks can be compiled at a single command. The generated document will automatically use the outputs from unit tests and module outputs from the latest runs, producing a file that visualizes additional checks of the system.  Thus, the validation document provide a more thorough test of the pipeline. We require agreement with a relative accuracy of $10^{-6}$ and absolute accuracy of $10^{-12}$ by default in the validation document, which maybe relaxed if necessary. In Fig.~\ref{fig:y1analysis_validation}, we show an excerpt of the document comparing predictions for cluster observable in the first year's of DES observation data (DES Y1), as an example of what to expect in this document. The output from our pipeline and the fiducial DES Y1 analysis pipeline, gives highly consistent predictions for the same set of cosmological and nuisance parameters.

\subsection{Instances of unit tests and physics validation}

We stress that both unit tests and physics validation are important to ensure the successful development and configuration of the pipeline. For example, the unit test of the 2-d interpolator caught an array index switch; finding this in a physics validation plot would be difficult. On the other hand, physics validation plots have caught errors that we were not able to uncover in unit tests. We illustrate this aspect with our experience of catching a mis-match in the installation of the pipeline. 
\begin{enumerate}
    \item Our analyses need to account for the effect of massive neutrinos in the galaxy cluster mass function, thus we developed modifications to the power spectrum implementation in the \textsc{CAMB} module in CosmoSIS which only uses the dark and baryonic matter components -- excluding the neutrino component -- of the power spectrum to compute the halo mass function. This is suggested in \cite{2013JCAP...12..012C} as massive neutrinos do not cluster at small scales. Because the modifications are carried out by a CosmoSIS module that is not part of the package, we attempted to establish a  branch in the CosmoSIS repository that contains the changes to be used in our development.
    \item The development of our package continued in the CosmoSIS branch. A few months later, we onboarded a new member to the team to use and participate in the development. The new member installed the software and ran the initial tests and validations.
    \item All unit tests pass with no issues.
    \item Upon running the physics-validation files in \textsc{R}, the predictions for the DES Y1 data vectors do not match comparison values. During the investigation, it was also discovered that the halo mass function output from the CosmoSIS pipeline do not match previous values. 
    \item We realized that our earlier modifications to the \textsc{CAMB} \citep{Lewis_2000, 2012JCAP...04..027H} module \footnotetext{https://camb.info}\footnotetext{ https://bitbucket.org/joezuntz/cosmosis/wiki/default\_modules/camb\_Jan15} a few months earlier have not been committed to the CosmoSIS repository. We promptly committed the branch and requested the CosmoSIS development team for approval, which since then has been merged into CosmoSIS.
\end{enumerate}

In this case, the success of our pipeline depends on successfully capturing the dependences  which are also undergoing constant developments and updates. Given that our modification of the \textsc{CAMB} module in CosmoSIS is not part of this package, those differences were not caught by unit tests. The separate validation testing which captures individual physics tests makes it possible to trace and debug this dependence.


\section{Analysis Cases}\label{sec:analysis}

\subsection{DES Y1 cluster Cosmology ``Mock'' Data}

\begin{figure*}
\label{lst:y1_analysis}
\begin{cpplisting}
std::vector<double>
y1_analysis_mor::operator()(double lo, double zt, double lnM) const
{

  std::vector<double> results(2 * zo_low_.size());
  double mass = std::exp(lnM);
  double common_term = (*mor)(lo, lnM, zt) *
     (*dv_do_dz)(zt) * (*hmf)(lnM, zt) * (*omega_z)(zt);
     
  // Number counts first in the returned results, 
  // followed by the masses
  for (std::size_t i = 0; i != zo_low_.size(); i++) {
    results[i] = (*int_zo_zt)(zo_low_[i], zo_high_[i], zt) * common_term;
    results[i + zo_low_.size()] = mass * results[i];
  }
  return results;
}
\end{cpplisting}
\caption{Snippet of the module operator function implementing the integrand in Equation 1. This operator function is part of the integration class, described in Section 4, implementing the DES Year 1 analysis. It makes use of the models described in Section 4 as well: for example (\* omega\_z)() is a call to the operator that implements the footprint function, similar to the one shown in Figure~\ref{lst:omega_sdss}. }
\label{lst:y1_analysis}
\end{figure*}

\begin{figure*}
\includegraphics[width=180mm]{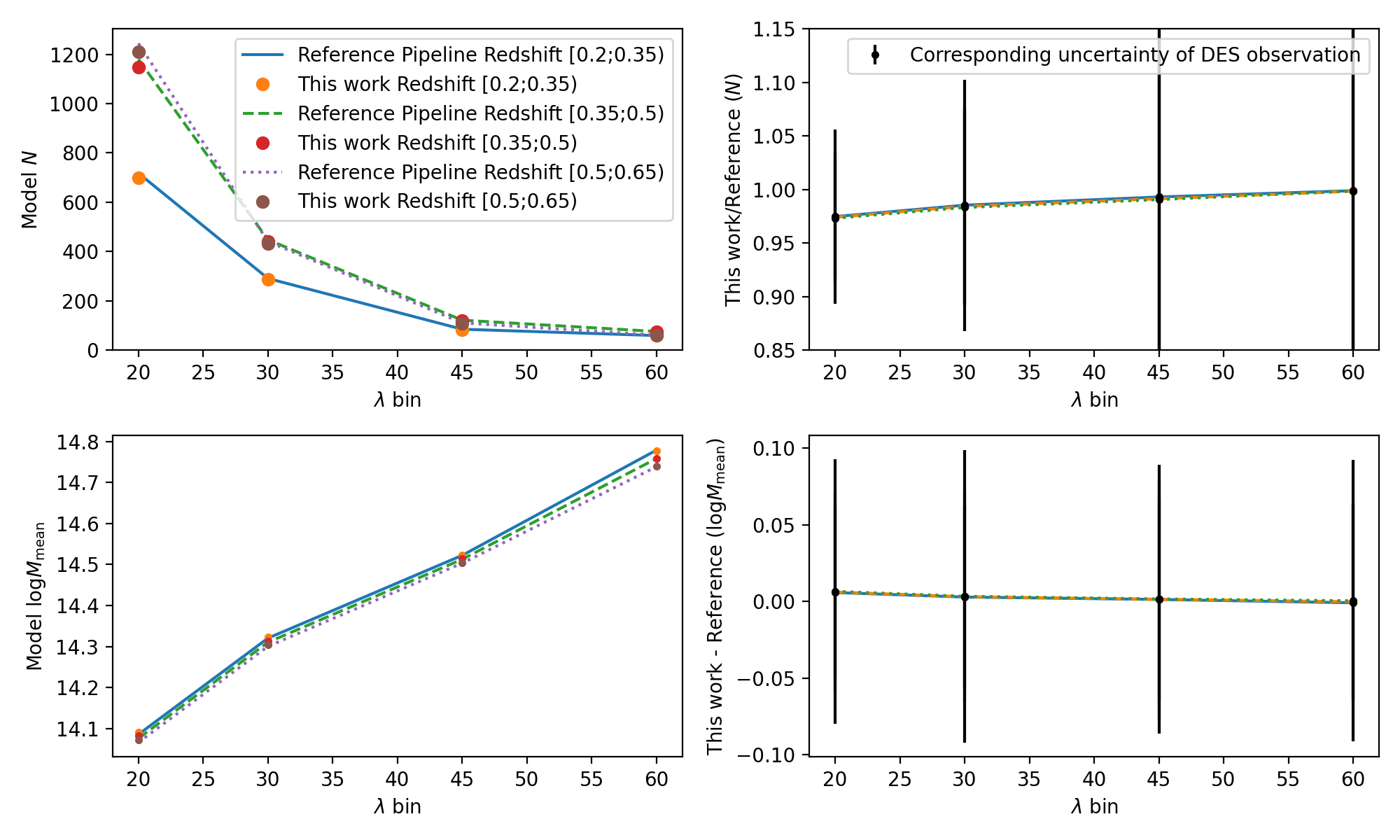}
\caption{Comparison between the theoretical predictions from this pipeline, and the reference pipeline used in DESY1 fiducial analyses. The upper panels show the predictions (left figure) of cluster number counts in three redshift ranges (see legend) and four richness ranges (x-axis), and the relative difference between the two pipelines (right figure). The lower panels show the predictions of clusters' average masses in the redshift/richness ranges (left figure) and the relative differences between the two pipelines (right figure). Note that the difference comparisons are highly consistent in the right figures, that the lines representing different redshift bins overlap.}\label{fig:y1_analysis_compar}
\end{figure*}

\begin{figure}
  \centering
   \includegraphics[width=90mm]{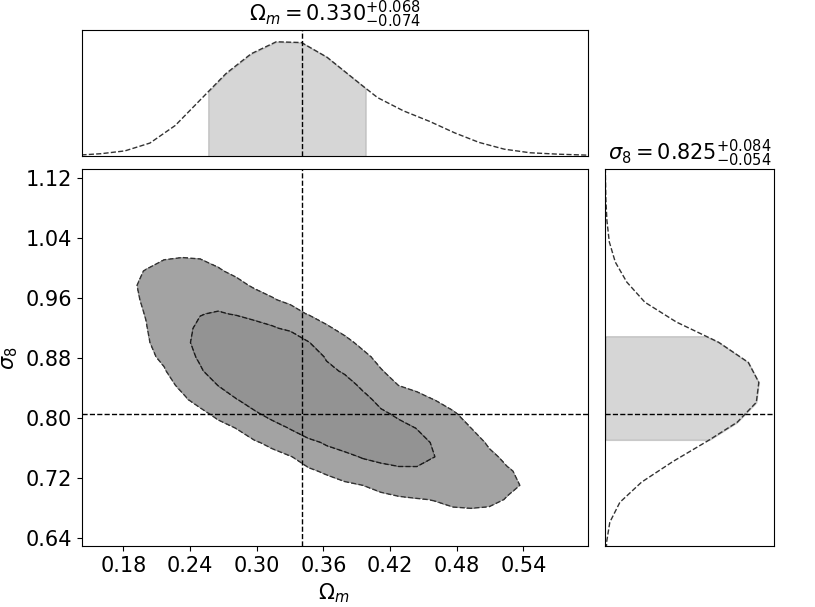}
  \caption{$\Omega_\mathrm{m} -\sigma_8$ posteriors derived from DES Y1  ``Mock'' data. The ``Mock''data are generated with $\Omega_\mathrm{m} =  0.3406$, and $\sigma_8 =0.8049$. Our pipeline successfully recovers the truth parameters with the 1$\sigma$ level. }
  \label{fig:mock_data_cosmos}  
\end{figure}

In this section, we demonstrate the efficacy of this pipeline by implementing a DES Year 1 cluster abundance cosmology analysis \cite{2020PhRvD.102b3509A} and compute prediction data vectors from the pipeline. In this demonstration, we model the number of clusters, and their average masses in richness/redshift subsamples. 

Specifically, we use the new pipeline to derive theoretical predictions of the cluster counts and average masses described by \autoref{eq:1}. As discussed in Section 4.3 and 4.4, we use the class template to implement the integrals of \autoref{eq:1} in CosmoSIS.  Figure~\ref{lst:y1_analysis} shows a code snippet implementing the integrands in those equations, which is a member function of the physics module class that will be used to calculate the integrals. 
Notably, we did not adopt a projection model \citep{2019MNRAS.482..490C} in the richness–mass relation in this analysis.  The projection model implementation does not significantly increase the computing time for a DES Y1-like analysis (Equation~\ref{eq:1}) as discussed in the next subsection, which is dominated by the running time of other cosmological modules outside of this development such as CAMB, that can take several seconds to run. 
Conversely, our modules dominate the running time when calculating cluster shear profiles in Equation~\ref{eq:shear-profile-BREAKDOWN},  which are much slower to calculate and can takes more than 100s of seconds using CPU machines; in this case, skipping the projection model greatly improves the running speed.

We compare the predictions of number counts and masses computed from our pipeline and the DES fiducial pipeline using one same set of cosmology and richness-mass relation parameters. Those comparisons are done five times with five sets of  parameters, and the predictions are all highly consistent. Figure ~\ref{fig:y1_analysis_compar}  shows the prediction data points for one set of the parameters (with $\Omega_\mathrm{m}= 0.3406$, $\sigma_8= 0.8049$): the number count predictions from the two pipelines agree within 3\% in the lowest richness bin, or within 1\% in the highest richness bin, and the mass predictions agree within 1\%. The prediction differences between the two pipelines are negligible compared to the measurement uncertainties.

To test the capability of our pipeline to recover unbiased parameter posteriors, we further use the prediction data vector as a ``Mock'' observational data vector (from the DES fiducial Year 1 pipeline), and derive the cosmological constraints for this ``Mock'' data set. Because the ``Mock'' observations are derived with the same  theoretical modelling code in this pipeline, we expect to recover the cosmological and richness-mass relation parameters that was used to derive those ``Mock'' data. To derive the posterior cosmological constraints, we set up the likelihood calculation: the integral output from the modules, which are the predictions for the cluster counts and average masses, are passed to a separate CosmoSIS likelihood module, as described by equation~\ref{eq:likelihood} in Section 3.  The likelihood is assumed to be a multi-dimensional Gaussian distribution with a covariance matrix (as described in \cite{2020PhRvD.102b3509A}).

The posterior parameter distributions are sampled with the ``polychord'' sampler implemented in CosmoSIS, as recommended in \citep[][private communications with the DES TCP Working Group]{2022arXiv220208233L} which is more accurate compared to the EMCEE sampler \citep{2013ascl.soft03002F} used in \cite{2020PhRvD.102b3509A} at sampling a large parameter space with the same number of samples. We vary and sample a total of 12 parameters (same with \cite{2020PhRvD.102b3509A} but setting both $\Omega_b$ and $\Omega_{\nu}$ to be 0), including cosmological parameters, cluster scaling relation parameters and nuisance parameters.
Finally, \autoref{fig:mock_data_cosmos} shows the posterior constraints on the  $\Omega_m$ and $\sigma_8$ parameters. For all of the 12 cosmological and richness-masss relation parameters, our pipeline has successfully recovered the values used to create the ``Mock" observations, within the $1\sigma$ level.

\subsection{DES Y1 cluster  cosmology constraints}

\begin{figure}
  \centering
   \includegraphics[width=90mm]{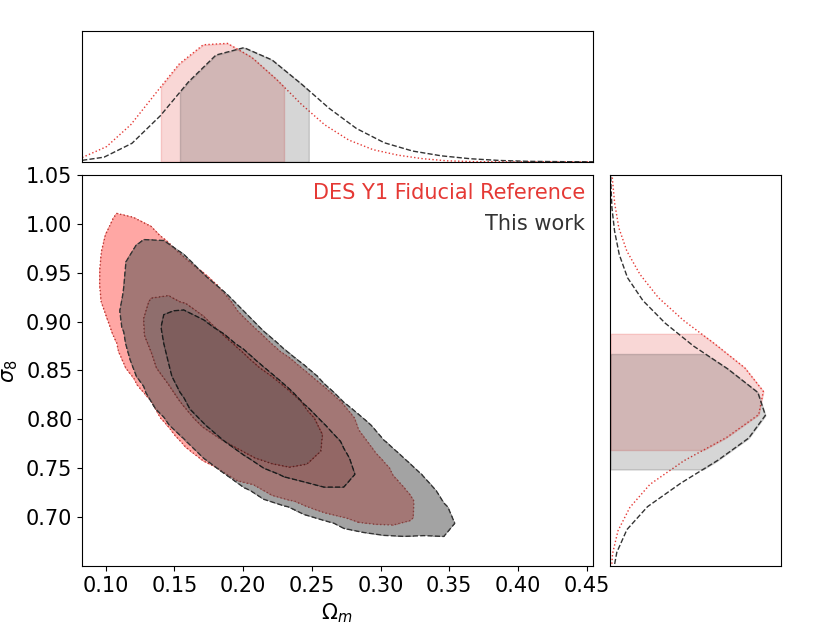}
  \caption{$\Omega_\mathrm{m}-\sigma_8$ posteriors from the DES Y1 cluster abundance cosmology analysis reanalyzed with a polychord sampler (red dotted line) and the result from the pipeline described in this paper (black dashed line). From DES Y1: $\Omega_\mathrm{m}$'s constraint is $0.191\pm0.043$, while $\sigma_8$'s constraint is $0.834\pm0.059$. From this work: $\Omega_\mathrm{m}$'s constraint is $0.210\pm0.046$, while $S_8$'s constraint is $0.813\pm0.057$. The results from the two pipelines are consistent within $0.5\sigma_8$ for both $\Omega_\mathrm{m}$ and $\sigma_8$.}
  \label{fig:y1analysis_comparison}  
\end{figure}

Based on the success with the DES Y1 ``Mock'' observations, we further use the pipeline to reanalyze the DES Year 1 data vector. We use the same prediction modules described in the previous subsection, but this time, using the real DES Year 1 data vector in the likelihood instead of the ``Mock'' observations. We sample a total f 14 cosmological and richness-mass parameters as presented in \cite{2020PhRvD.102b3509A}, using the same prior distribution for those parameters. 

\autoref{fig:y1analysis_comparison} shows the posterior constraints on the $\Omega_m$ and $\sigma_8$ parameters from our re-analysis. 
For comparison, we also show the $\Omega_m$ and $\sigma_8$ constraints from the fiducial DES analysis pipeline of the same data vectors but using the polychord sampler. The results are consistent within 0.5$\sigma$ of the fidual DES analysis despite the differences in the software pipelines. We note that we have tried different prior setups for the parameters, and found that in one case, the differences can be further reduced to within 0.3$\sigma$, but the posteriors presented here are based on the closest implementation of the DES fiducial analysis (excluding projection effect in the richness-mass relation). Because our pipeline is an independent implementation from the DES fiducial analysis, using different programming languages (C++ {\it VS} Fortran), integration algorithms, and pipeline design, the consistency in the results between the two pipelines also highlight the robustness of the cosmological constraints from the DES fiducial analysis. 

 The CosmoSIS module we implemented for this re-analysis, which makes theoretical predictions of the cluster counts and average masses in a variety of integration volumes using \autoref{eq:1} is fast. Using a single processor on a Macbook pro (2GHz intel core i5), the calculation for all of the 12 richness and redshift bins (corresponding integration volumes in our pipeline) listed in the DES Y1 analysis takes less than 0.1 second for one set of cosmological and nuisance parameters. Including the projection effect model increases the computing time to about 1.9 seconds. Both impelmentations are significantly faster than the other CosmoSIS modules needed for the DES Y1 analysis, such as the CosmoSIS CAMB module (which can take $\sim$ 4 seconds). Notably, the speed of those computations change with the values of cosmological and nuisance parameters, as the integrations in \autoref{eq:1} may be slow to converge if the parameters are far away from a likely range (but necessary to sample in the MCMC process).

\subsection{DES Y1 shear profile modeling}

\begin{figure*}
  \centering
  	\includegraphics[width=180mm]
   {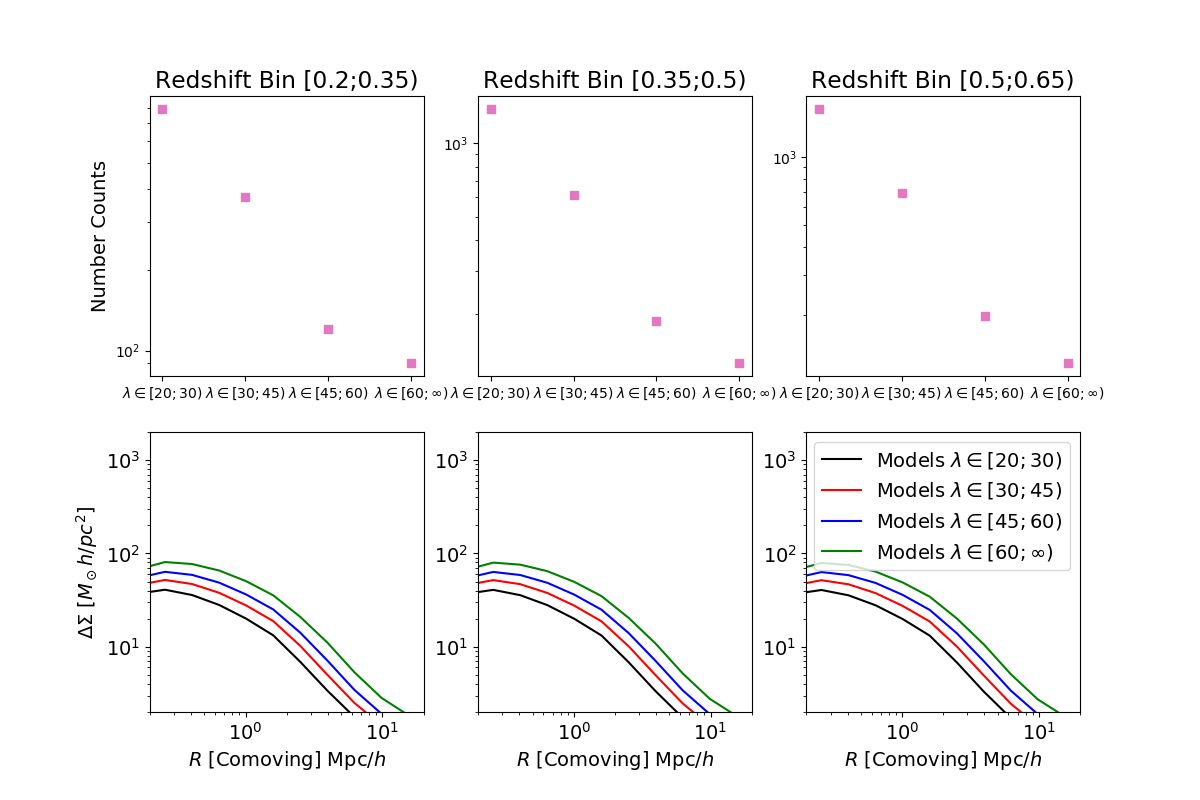}
  \caption{Predictions from our pipeline for cluster number counts (upper panel) and their weak lensing measurement signals (lower panel) in multiple redshift (each subplot) and richness ranges (each data point/line) for one set of cosmological and nuisance parameters (not at ``best-fit"). In cosmological analysis, those predictions will be compared to constrain cosmological parameters. We defer that analysis to a future study. }
  \label{fig:y1analysis_deltasigma}
\end{figure*}

The cluster masses that we used as an observable in DES Y1 cluster abundance analysis are further derived from the cluster-weak lensing shear measurements \cite{2019MNRAS.482.1352M}. Given an efficient method, we can directly model cosmology from cluster number counts and cluster weak-lensing measurements, bypassing the step to derive masses. Here, we further demonstrate that the pipeline can be used to compute the predictions of the DES Y1 weak-lensing observables based on Equations \ref{eq:shear-profile-BREAKDOWN} to \ref{eq:DS-breakdown}. Using a set of cosmological and nuisance parameters from the posterior distribution in the previous section (but not at ``best-fit"), and assuming a concentration parameter, we show those predictions in Figure~\ref{fig:y1analysis_deltasigma}. 

Ideally, we will also constrain cosmological parameters through modeling the measurements, and sampling using MCMC. However, a fully accurate pipeline will require a good understanding of the selection effects \citep{2020PhRvD.102b3509A} as well as a robust procedure of testing on simulations \citep{2021arXiv210513547D, 2021MNRAS.502.4093T} and a blinding scheme \citep{2020MNRAS.494.4454M}. We defer this analysis to an ongoing study. In addition, the analysis pipeline developed for this purpose is also being implemented for GPU computing clusters to speed up the analysis. We will discuss those implementations in other publications.


\section{Discussion}

In this paper, we describe a software development effort to  deliver theoretical modeling of galaxy cluster cosmology observables implemented in  CosmoSIS  for cosmological likelihood calculations. Given the set of astrophysical and cosmological factors to be taken into account and thus the challenging computing demands, this sofware package is developed to optimize speed and accuracy. Specifically, we address several challenges in the development efforts. 

First, the prediction of the cluster cosmological observables depend on a few models for different physics elements. Each of the models, for example the cluster mass-observable relation, halo mass function etc., is still under active development, and subject to active updates. In addition, it is often desirable to explore several different forms of those models to assess the robustness of the analysis due to different physical assumptions. Two examples are the models for the mass-observable relations and the halo mass function.
To make it easy to change a model implementation, we make the models modular, that one model implementation can be replaced by another that shares use of the same set of physical parameters. Each model is represented as a C++ class, which is \emph{callable} by implementing the function call operator. 

Second, the predictions can be implemented as numerical integrals as shown in Section~\ref{s:theory}. The dimensions of the integrals are often larger than two, and in our case, reaches as many as eight. Common integration tools such as those implemented in Python, or GSL, have only implemented algorithms to handle 1-D or 2-D integrals. 
The dimension of those integrals can become even larger for future analyses if marginalizing over parameters representing additional systematic effects.  We make use of the algorithms described in~\cite{Berntsen1991AnAA}, and implemented in the CUBA library~\cite{HAHN200578} in our development.

Thirdly, the integrals mentioned above need to be evaluated for each sample of cosmological and nuisance parameters. If using MCMC sampling, those integrals often need to be iterated $\sim 10^{6}$ times to evaluate the posterior distribution of cosmological parameters. In order to finish the sampling in a reasonable timescale (ideally in a matter of days), the integrals need to be performed very quickly. We  implement all our models and our integration routines in C++:
it provides the best combination of expressiveness and performance available. 
We also implement procedures to ensure the reproducibility of the code, achieved through testing. 

In this paper, we demonstrate the effectiveness of this software package by reproducing the cluster cosmology analysis based on the public DES Year 1 data sets of cluster number counts and weak-lensing masses. This analysis has also served as an independent check of the DES Y1 fiducial analysis, verifying the robustness of the previous analysis. In the future, we expect to further adapt the software for GPU resources \citep{Sakiotis2021PAGANIAP}. Using this package, we will further extend the DES Year 1 analysis by forward modeling cluster number counts and weak lensing radial signals. The analysis is currently under-way and will be described in a future publication. 



\section{Acknowledgement}

MESP is funded by the Deutsche Forschungsgemeinschaft (DFG, German Research Foundation) under Germany's Excellence Strategy – EXC 2121 "Quantum Universe" – 390833306.

Funding for the DES Projects has been provided by the U.S. Department of Energy, the U.S. National Science Foundation, the Ministry of Science and Education of Spain, 
the Science and Technology Facilities Council of the United Kingdom, the Higher Education Funding Council for England, the National Center for Supercomputing 
Applications at the University of Illinois at Urbana-Champaign, the Kavli Institute of Cosmological Physics at the University of Chicago, 
the Center for Cosmology and Astro-Particle Physics at the Ohio State University,
the Mitchell Institute for Fundamental Physics and Astronomy at Texas A\&M University, Financiadora de Estudos e Projetos, 
Funda{\c c}{\~a}o Carlos Chagas Filho de Amparo {\`a} Pesquisa do Estado do Rio de Janeiro, Conselho Nacional de Desenvolvimento Cient{\'i}fico e Tecnol{\'o}gico and 
the Minist{\'e}rio da Ci{\^e}ncia, Tecnologia e Inova{\c c}{\~a}o, the Deutsche Forschungsgemeinschaft and the Collaborating Institutions in the Dark Energy Survey. 

The Collaborating Institutions are Argonne National Laboratory, the University of California at Santa Cruz, the University of Cambridge, Centro de Investigaciones Energ{\'e}ticas, 
Medioambientales y Tecnol{\'o}gicas-Madrid, the University of Chicago, University College London, the DES-Brazil Consortium, the University of Edinburgh, 
the Eidgen{\"o}ssische Technische Hochschule (ETH) Z{\"u}rich, 
Fermi National Accelerator Laboratory, the University of Illinois at Urbana-Champaign, the Institut de Ci{\`e}ncies de l'Espai (IEEC/CSIC), 
the Institut de F{\'i}sica d'Altes Energies, Lawrence Berkeley National Laboratory, the Ludwig-Maximilians Universit{\"a}t M{\"u}nchen and the associated Excellence Cluster Universe, 
the University of Michigan, NSF's NOIRLab, the University of Nottingham, The Ohio State University, the University of Pennsylvania, the University of Portsmouth, 
SLAC National Accelerator Laboratory, Stanford University, the University of Sussex, Texas A\&M University, and the OzDES Membership Consortium.

Based in part on observations at Cerro Tololo Inter-American Observatory at NSF's NOIRLab (NOIRLab Prop. ID 2012B-0001; PI: J. Frieman), which is managed by the Association of Universities for Research in Astronomy (AURA) under a cooperative agreement with the National Science Foundation.

The DES data management system is supported by the National Science Foundation under Grant Numbers AST-1138766 and AST-1536171.
The DES participants from Spanish institutions are partially supported by MICINN under grants ESP2017-89838, PGC2018-094773, PGC2018-102021, SEV-2016-0588, SEV-2016-0597, and MDM-2015-0509, some of which include ERDF funds from the European Union. IFAE is partially funded by the CERCA program of the Generalitat de Catalunya.
Research leading to these results has received funding from the European Research
Council under the European Union's Seventh Framework Program (FP7/2007-2013) including ERC grant agreements 240672, 291329, and 306478.
We  acknowledge support from the Brazilian Instituto Nacional de Ci\^encia
e Tecnologia (INCT) do e-Universo (CNPq grant 465376/2014-2).

This manuscript has been authored by Fermi Research Alliance, LLC under Contract No. DE-AC02-07CH11359 with the U.S. Department of Energy, Office of Science, Office of High Energy Physics.

\bibliography{mybibfile}

\appendix

\section{Institutions}

{
\scriptsize
$^{1}$ Laborat\'orio Interinstitucional de e-Astronomia - LIneA, Rua Gal. Jos\'e Cristino 77, Rio de Janeiro, RJ - 20921-400, Brazil\\
$^{2}$ Department of Physics, University of Michigan, Ann Arbor, MI 48109, USA\\
$^{3}$ Fermi National Accelerator Laboratory, P. O. Box 500, Batavia, IL 60510, USA\\
$^{4}$ Institute of Cosmology and Gravitation, University of Portsmouth, Portsmouth, PO1 3FX, UK\\
$^{5}$ University Observatory, Faculty of Physics, Ludwig-Maximilians-Universit\"at, Scheinerstr. 1, 81679 Munich, Germany\\
$^{6}$ Department of Physics \& Astronomy, University College London, Gower Street, London, WC1E 6BT, UK\\
$^{7}$ Universidad de La Laguna, Dpto. Astrofisica, E-38206 La Laguna, Tenerife, Spain\\
$^{8}$ Instituto de Astrof\'isica de Canarias, E-38205 La Laguna, Tenerife, Spain\\
$^{9}$ Kavli Institute for Cosmological Physics, University of Chicago, Chicago, IL 60637, USA\\
$^{10}$ Department of Astronomy and Astrophysics, University of Chicago, Chicago, IL 60637, USA\\
$^{11}$ Astronomy Unit, Department of Physics, University of Trieste, via Tiepolo 11, I-34131 Trieste, Italy\\
$^{12}$ INAF-Osservatorio Astronomico di Trieste, via G. B. Tiepolo 11, I-34143 Trieste, Italy\\
$^{13}$ Institute for Fundamental Physics of the Universe, Via Beirut 2, 34014 Trieste, Italy\\
$^{14}$ School of Mathematics and Physics, University of Queensland,  Brisbane, QLD 4072, Australia\\
$^{15}$ Centro de Investigaciones Energ\'eticas, Medioambientales y Tecnol\'ogicas (CIEMAT), Madrid, Spain\\
$^{16}$ Jet Propulsion Laboratory, California Institute of Technology, 4800 Oak Grove Dr., Pasadena, CA 91109, USA\\
$^{17}$ Institute of Theoretical Astrophysics, University of Oslo. P.O. Box 1029 Blindern, NO-0315 Oslo, Norway\\
$^{18}$ SLAC National Accelerator Laboratory, Menlo Park, CA 94025, USA\\
$^{19}$ Center for Astrophysical Surveys, National Center for Supercomputing Applications, 1205 West Clark St., Urbana, IL 61801, USA\\
$^{20}$ Department of Physics and Astronomy, University of Pennsylvania, Philadelphia, PA 19104, USA\\
$^{21}$ Institut de F\'{\i}sica d'Altes Energies (IFAE), The Barcelona Institute of Science and Technology, Campus UAB, 08193 Bellaterra (Barcelona) Spain\\
$^{22}$ Department of Astronomy, University of Illinois at Urbana-Champaign, 1002 W. Green Street, Urbana, IL 61801, USA\\
$^{23}$ Santa Cruz Institute for Particle Physics, Santa Cruz, CA 95064, USA\\
$^{24}$ Center for Cosmology and Astro-Particle Physics, The Ohio State University, Columbus, OH 43210, USA\\
$^{25}$ Department of Physics, The Ohio State University, Columbus, OH 43210, USA\\
$^{26}$ Center for Astrophysics $\vert$ Harvard \& Smithsonian, 60 Garden Street, Cambridge, MA 02138, USA\\
$^{27}$ Department of Physics, University of Arizona, Tucson, AZ 85721, USA\\
$^{28}$ Lowell Observatory, 1400 Mars Hill Rd, Flagstaff, AZ 86001, USA\\
$^{29}$ Australian Astronomical Optics, Macquarie University, North Ryde, NSW 2113, Australia\\
$^{30}$ Department of Physics, Carnegie Mellon University, Pittsburgh, Pennsylvania 15312, USA\\
$^{31}$ George P. and Cynthia Woods Mitchell Institute for Fundamental Physics and Astronomy, and Department of Physics and Astronomy, Texas A\&M University, College Station, TX 77843,  USA\\
$^{32}$ Jodrell Bank Center for Astrophysics, School of Physics and Astronomy, University of Manchester, Oxford Road, Manchester, M13 9PL, UK\\
$^{33}$ Instituci\'o Catalana de Recerca i Estudis Avan\c{c}ats, E-08010 Barcelona, Spain\\
$^{34}$ Hamburger Sternwarte, Universit\"{a}t Hamburg, Gojenbergsweg 112, 21029 Hamburg, Germany\\
$^{35}$ Observat\'orio Nacional, Rua Gal. Jos\'e Cristino 77, Rio de Janeiro, RJ - 20921-400, Brazil\\
$^{36}$ Kavli Institute for Particle Astrophysics \& Cosmology, P. O. Box 2450, Stanford University, Stanford, CA 94305, USA\\
$^{37}$ Department of Physics and Astronomy, Pevensey Building, University of Sussex, Brighton, BN1 9QH, UK\\
$^{38}$ Department of Physics and Astronomy, Stony Brook University, Stony Brook, NY 11794, USA\\
$^{39}$ School of Physics and Astronomy, University of Southampton,  Southampton, SO17 1BJ, UK\\
$^{40}$ Computer Science and Mathematics Division, Oak Ridge National Laboratory, Oak Ridge, TN 37831\\
$^{41}$ \\
$^{42}$ Universit\"ats-Sternwarte, Fakult\"at f\"ur Physik, Ludwig-Maximilians Universit\"at M\"unchen, Scheinerstr. 1, 81679 M\"unchen, Germany\\
$^{43}$ Max Planck Institute for Extraterrestrial Physics, Giessenbachstrasse, 85748 Garching, Germany\\
$^{44}$ Department of Physics, Boise State University, Boise, ID 83725, USA\\
$^{45}$ NSF's National Optical-Infrared Astronomy Research Laboratory, 950 N. Cherry Ave., Tucson, AZ, 85719, USA\\
$^{46}$ Department of Physics, Duke University Durham, NC 27708, USA\\
}

\end{document}